\documentclass[prl,twocolumn]{revtex4}
\usepackage{graphicx}% Include figure files
\usepackage{dcolumn}% Align table columns on decimal point
\usepackage{bm}% bold math
\newcommand{\figwidth}{3. in}
\begin{document}
\title{\bf Possible observation of  ``string excitations'' of a hole
in a quantum antiferromagnet}
\author{Efstratios Manousakis}
\affiliation{Department of Physics, Florida State University, 
Tallahassee, FL 32306-4350 \\
and Department of  Physics, University of Athens, Greece.}  
\date{\today}
\begin{abstract}
%When a hole is created in a quantum antiferromagnet 
%it becomes a mobile well-defined quasiparticle with a dispersion 
%which has its minimum at $(\pi/2,\pi/2)$. For short time scales, however,
%the hole feels a linearly rising potential as it attempts to move away
%from the most  favorable position.
We argue that recently reported high resolution
angle-resolved photoelectron spectra from
cuprates, where an anomalous high-energy dispersion
was identified, reveal the internal structure of the hole quasiparticle
in quantum antiferromagnets and more importantly it is evidence 
for the existence of ``string-excitations'' which validate early predictions
based on the $t-J$ model. Their energy-momentum despersion as well
as the manner in which the spectral weight is transfered to higher
energy string excitations as well as the vanishing of the
quasiparticle spectral weight near the $\Gamma$ point, are all in 
agreement with predictions without adjusting any parameters.

\end{abstract}

\maketitle

%\underbar{Introduction:}
The study of the motion of a single hole in a quantum antiferromagnet is
of general theoretical and experimental importance not only because 
it might be pertinent to the mechanism of superconductivity in the 
cuprates, but also because it is relevant to the field of
quantum magnetism and strongly correlated systems, and has connections 
and analogies with
the problem of  impurity motion in antiferromagnets and
in quantum solids and quantum liquids. 

There is a solid body of angle-resolved photoelectron spectroscopy (ARPES)
studies\cite{damascelli} which reveal important features of the insulating, 
lightly doped, and overdoped cuprates. Here, we  focus first on ARPES studies
of the single hole dispersion in an undoped insulating antiferromagnetic
parent compound. Early such studies\cite{wells} demonstrated that 
there is a sharp well-defined quasiparticle-like peak in the spectral
function which as function of momentum defines a band with a minimum 
near $(\pi/2,\pi/2)$ and a characteristic bandwidth
approximately $2.2 J$, where $J$ is the antiferromagnetic coupling. 
These features had been 
predicted by a number of studies of the hole motion in a quantum 
antiferromagnet\cite{liu,shraiman,elser,KLR,marsiglio,martinez,boninsegni,dagottor} in its simplest conception where a hole hopping term 
is added to the Heisenberg antiferromagnetic Hamiltonian\cite{RMP},
the so-called $t-J$ model\cite{ZR}. 
A deficiency of the simple $t-J$ model is that for momentum near $(\pi,0)$ it 
gives\cite{liu}  a spectral function similar in shape and energy 
to that at $(\pi/2,\pi/2)$, while the 
 ARPES measurements\cite{wells} revealed that the quasiparticle peak
is broader near $(\pi,0)$ and the corresponding energy is
higher than that at $(\pi/2,\pi/2)$. 
This discrepancy can be removed by adding relatively small  
direct next-nearest-neighbor hopping terms ($t^{\prime}$ and 
$t^{\prime\prime})$ in the $t-J$ 
model\cite{kim,theory,nazarenco,ferrell,lee,wheatley,belincher,eder}.

However, in recent ARPES studies\cite{ronning,graf}, high resolution data, 
taken along the $(0,0)$ to $(\pi,\pi)$ cut, show an additional 
dispersive feature
at higher energies that merges with the above mentioned band
at lower energies. The main point of the present paper is to argue that these
high resolution ARPES studies have revealed the internal structure of the 
single-hole quasiparticle as well as the existence and the energy-momentum
dispersion of 
 ``string excitations''\cite{shraiman,elser,KLR,barnes,dagotto,liu}. 

\begin{figure}[ht] 
\vskip 0.2 in
\begin{center}
  \includegraphics[width=2.5 in]{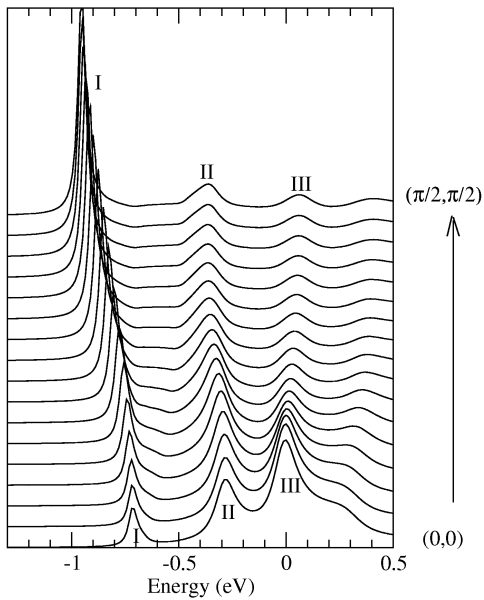}
\vskip 0.2 in
 \caption{\label{fig1} The spectral function of a single hole
in a quantum antiferromagnet for $J/t=0.3$ along the $(0,0)$ to 
$(\pi/2,\pi/2)$ direction as a function of energy $\omega$. The
Dyson's equation has been solved as in Ref.~\onlinecite{liu}
using Lorentzian broadening with a width $\eta=0.1 t$. While the width of the
first peak depends strongly on the value of $\eta$, as it is a
$\delta$ function peak, the other two (labeled II and III) and the 
higher peaks (not shown in the graph) remain unchanged
when we decrease the value of $\eta$. This point has been extensively
studied in  Ref.~\onlinecite{liu} and we have reproduced it here.}
\end{center}
\end{figure}

When a quantum hole is
created in a classical antiferromagnet (such as, the $t-J_z$ model), 
the hole stays bound to its 
``birth-site'' due to a ``string'' of overturned spins produced by the hole 
motion in its attempt to compromise its uncertainty in momentum 
by allowing for some position uncertainty. In a quasi-continuum 
picture, the hole is trapped by a linearly rising potential
characterized by energy eigenstates, the so-called Airy 
functions\cite{shraiman,KLR,elser,liu},
with energies $E_n/t = \epsilon_n + a_n (J_z/t)^{2/3}$,
and the average length of the string scales as $(t/J_z)^{1/3}$
(here $t$ is the hole-hopping matrix element and $J_z$ is the coupling 
along the $z$ direction in spin-space). By turning on quantum 
spin-fluctuations, i.e., in the case of a mobile hole in a
quantum antiferromagnet, the hole becomes  a well-defined delocalized 
quasiparticle with the band minimum at $(\pi/2,\pi/2)$
and a bandwidth of the order of the antiferromagnetic spin-exchange
coupling. In Ref.~\onlinecite{liu}, the string-excitations 
were extensively studied
and it was found that they survive the turning on of quantum 
spin-fluctuations,  i.e., they give rise to {\it rather well-defined peaks} 
in the hole spectral function at higher energies. In addition, as it is 
shown in this paper (see Fig.~\ref{fig1} to be discussed
later), they are responsible for the transfer of the quasiparticle 
weight from the low energy minimum at $(\pi/2,\pi/2)$ to the
second and third string excitation as the $\Gamma$ point ($(0,0)$) 
is approached
and they are responsible for the observed vanishing 
of the quasiparticle weight near the $\Gamma$  point\cite{ronning}. 

Next, in this paper, we will argue that the recently published results of the
high resolution ARPES study or Ref.~\onlinecite{ronning} on the insulating 
cuprate $Ga_2CuO_2Cl_2$ provide strong evidence for the existence
of these string-excitations. Furthermore, they yield their energy-momentum 
dispersion in agreement with the predictions made starting from the 
$t-J$ model\cite{liu}. More generally these ARPES studies illuminate 
the role of string excitations in lightly doped quantum 
antiferromagnets and validate the theoretical
framework which predicted them\cite{shraiman,liu}.
 
%\underbar{Making the case:}
In order to make our arguments more convincing we will 
use the simpler $t-J$ model using the widely accepted
values of the parameters $J/t=0.3$ and $t=0.4 eV$\cite{kim,theory,nazarenco,ferrell,lee,wheatley,belincher,eder,boninsegni}.  
While, as we discussed there
is a need to introduce next-nearest-neighbor hopping terms
in order to reproduce the features of the quasiparticle
band near $(\pi,0)$, we will restrict our studies along
the $(0,0)$ to $(\pi,\pi)$ cut where these terms do not have a
significant effect. Therefore, we will use the pure $t-J$ model
with no free parameters, because the elegant aspects of the 
phenomenon that we try to convey can be described more clearly.

\begin{figure}[ht] 
\begin{center}
\includegraphics[width=\figwidth]{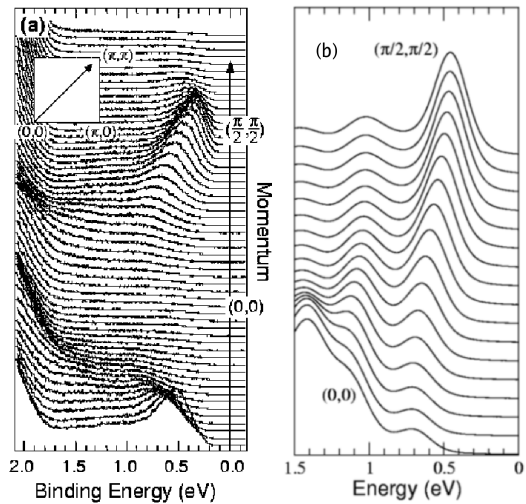}
 \caption{\label{fig2} 
Comparison between (a) the experimentally determined spectral function
with high resolution ARPES\cite{ronning}  and (b) that obtained
from the $t-J$ model\cite{liu} for $J/t=0.3$ and $t=0.4 eV$ along the path from
$(0,0) \to (\pi/2,\pi/2)$. The reference energy has been shifted by a 
constant.}
\end{center}
\end{figure}

In Fig.~\ref{fig1} the spectral function of the $t-J$ model
is presented for $J/t=0.3$ along the direction $(0,0) \to (\pi/2,\pi/2)$
calculated with the same approach as in Ref.~\onlinecite{liu}.
There is the main quasiparticle peak labeled I at low
frequencies and at least two more visible peaks labeled II and III
at higher frequencies. These peaks correspond to higher energy eigenstates
of the simple 2D quantum ``yo-yo'' problem (a particle in a
linearly rising potential) with the additional
complication due to the fact that there is the quantum spin
exchange term and the physics of this will be discussed later in
this paper. One of the important aspects of the graph is the transfer
of the spectral weight from the lowest energy peak 
(peak I) to the higher energy peak as the momentum 
changes from $(\pi/2,\pi/2)$ to $(0,0)$. The physical
explanation for the transfer of weight will be also given below
after we make our case by comparing with the experiment.
The second important aspect of the graph is that the spectral
weight of the lowest energy peak nearly vanishes at the 
$\Gamma$  point of the Brillouin zone.

\begin{figure}[ht] 
\begin{center}
\includegraphics[width=\figwidth]{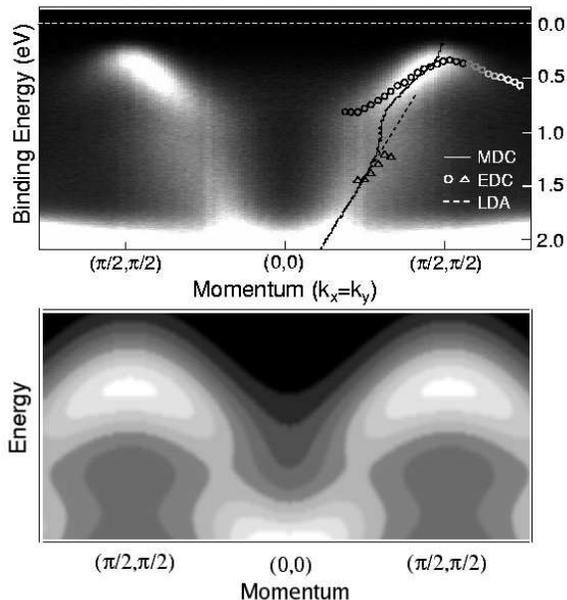}
 \caption{\label{fig3} 
Comparison of the experimentally observed intensity 
plot reported in Fig.~2 of Ref.~\onlinecite{ronning} (top) with
that obtained from the $t-J$ model (bottom). The vertical energy scale
in the theoretical curve (bottom) is similar to the experimental 
scale (top), i.e., about $1.5 eV$. Notice the
intense peaks at $(\pi/2,\pi/2)$ and near $(0,0)$ in both theoretical 
and experimental intensity plots.
At momentum $(0,0)$ the spectral weight has been
transfered to higher energy ``string'' states (II and III) as also
seen in Fig.~\ref{fig1} and Fig.~\ref{fig2}. This
gradual transfer manifests itself as a more luminous path
connecting the bright peaks at $(\pi/2,\pi/2)$ and $(0,0)$. }
\end{center}
\end{figure}
\begin{figure}[ht] 
\begin{center}
\includegraphics[width=2.5 in]{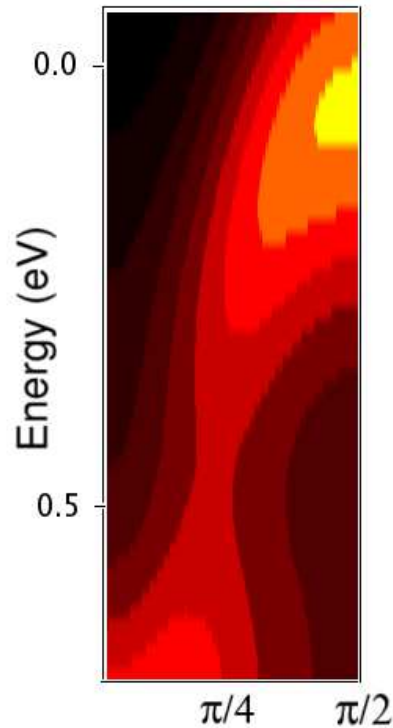}
 \caption{\label{fig4} 
The theoretical color coded intensity plot obtained from the $t-J$ model
using the widely acceptable values for the parameters $J/t=0.3$ and $t=0.4 eV$.
It should be compared with the experimentally reported in Fig.~1 
of Ref.~\onlinecite{graf}.}
\end{center}
\end{figure}
%\begin{figure}[ht] 
%\begin{center}
%\includegraphics[width=3.0 in]{alongxboth.eps}
% \caption{\label{fig4} 
%}
%\end{center}
%\end{figure}
 
In Fig.~\ref{fig2} and Fig.~\ref{fig3} a comparison is made between 
the results of high resolution ARPES\cite{ronning} as have been
presented in Fig.~1 and Fig.~2 of  Ref.~\onlinecite{ronning}. The
theoretical results have been further broadened with a
Gaussian broadening function, namely $A_b({\bf k},\omega)
= \int A({\bf k},\omega') G(\omega-\omega') d\omega'$, where
$G(\omega-\omega') = exp(-(\omega-\omega')^2/\sigma^2)$, using 
$\sigma=0.125 eV$. Very similar results have been obtained if
we use a value of $\eta=0.125 eV$ in the initial propagator 
(i.e., Lorenzian broadening) when we solve the Dyson's equation.
 The same amount of broadening is also necessary to broaden the theoretical
sharp quasiparticle peak so that its line-width is the same as the
experimental one. This amount of broadening is very close to that
used in Ref.~\onlinecite{kim} to compare the ARPES peak to the
quasiparticle peak given by the $t-J$ model and its extensions.
%There is a tendency in the microscopic calculation to 
%give sharper peaks and this broadening is done in order to smooth out the
%peaks.
%The $t-J$ model does not include the coupling of the
%hole and spins to a number of other excitations which are present in 
%the cuprate material and might be 
%expected to cause broadening. 
%In addition, in specific regions the 
%luminosity reported by the ARPES study  is low and therefore 
%the resolution in such regions is expected to be low. 
Notice, in Fig.~\ref{fig2} the transfer of weight  from the lowest energy peak
(most prominent at $(\pi/2,\pi/2)$) to another higher
energy peak which forms as the momentum $(0,0)$ is approached starting from
$(\pi/2,\pi/2)$.  This seems to be the case in the most accurate ARPES data
available from the insulator\cite{ronning} and it becomes clearer in 
the following figure.

In Fig.~\ref{fig3} we compare the experimentally obtained\cite{ronning} 
intensity plot (top) to that obtained from the $t-J$ model
for the same parameter values as those of Fig.~\ref{fig1} and the
broadening procedure discussed above. Notice again that the gradual
transfer of the spectral weight from the lowest 
energy peak to the higher energy peaks
(mainly to the peaks labeled II and III  in Fig.~\ref{fig1})  
appears as an ``anomalous'' high energy dispersion due to 
broadening and limited resolution. 
In other words, what appears to be a high energy dispersive  curve
is the dispersion of the center of ``gravity'' of the peaks 
II and III as they become more and more luminous as the
value of momentum approaches $(0,0)$.

Most recently in a different ARPES study an anomalous
 dispersion and a second energy  scale at around $0.8 eV$ 
was reported\cite{graf}.  This energy
scale appears to be the center of ``gravity'' of the string excitation 
peaks II and  III (measured from the lowest energy state at 
$(\pi/2,\pi/2)$) in our calculation, as can be seen by comparing 
our Fig.~\ref{fig1} with Fig.~2 of Ref.~\onlinecite{graf}.
 Notice that, assuming that these features of the hole-band 
do not significantly change by doping,  when the lowest energy quasiparticle
states,  which correspond
to the lowest energy string states in the neighborhood of $(\pi/2,\pi/2)$, are 
occupied the higher lying string excitations will be probed by ARPES.
The color-coded intensity plot of Fig.~\ref{fig4}, excluding the bright
spots around $(\pi/2,\pi/2)$ (because at sufficient amount of doping the
states around $(\pi/2,\pi/2)$  should be inside the Fermi sea and the 
bright spots  should move outside the Fermi surface), agrees reasonably
well with the experimental intensity plots reported in 
Fig.~1 of Ref.~\onlinecite{graf}.
As can be clearly seen from these figures and the previous discussion,
the process of  spectral weight  transfer to the higher
string states is masked by low intensity and broadening and
shows up as an ``anomalous'' dispersion with
the center of the anomaly close to $(\pi/4,\pi/4)$ as in Fig.~1 of
Ref.~\onlinecite{graf}. It is surprising and perhaps revealing
that these features persist all the way up to the overdoped regime.
Therefore, it appears that the spin correlations
should be strongly antiferromagnetic even in the overdoped
cuprates. 

%\begin{figure}[ht] 
%\begin{center}
%  \includegraphics[width=\figwidth]{gnuplot.eps}
% \caption{\label{fig1} The spectral function of a single hole
%in a quantum antiferromagnet for $J/t=0.3$ along the $(0,0)$ to 
%$(\pi/2,\pi/2)$ direction
%as a function of $-\omega$ (in units of $t$).}
%\end{center}
%\end{figure}
%\underbar{The origin of the peaks}:
We can give a qualitative picture of the origin of the spectral function
peaks that correspond to the string excitations and
a qualitative explanation of the spectral weight shift. Let us first discuss
the single  hole motion in a classical antiferromagnet, the so-called
$t-J_z$ model.
In this case as the hole moves away from its ``birth'' site, it
displaces spins and, thus, it feels a potential which, as a function of the
length of the hole path from its ``birth'' site, is linear with slope 
equal to $J_z$. Therefore, in this limit
the hole is almost localized and, for relatively large values of $t/J_z$, 
a quasi-continuum picture could be used to qualitatively describe 
the hole motion, where the energy levels are given by the form 
$E_n/t = \epsilon_n + a_n (J_z/t)^{2/3}$ and the corresponding
wave-functions are the Airy functions. Once the Heisenberg spin-exchange
term $J_{\perp}/2(s^+_i s^-_j + s^-_i s^+_j)$ is turned on,
two ``string'' states of overturned spins with the ``birth'' site at the 
beginning of each string and the hole at the end, which only
differ by just two spins at the beginning of the string, 
can have significant overlap through this Heisenberg spin-exchange.
This non-zero overlap can give favorable (lowering) contribution
to the hole kinetic energy if the phase-difference between the 
amplitudes in the quasihole wavefunction
associated with these two related strings is $e^{i{\bf k}\cdot {\bf R}} = -1$.
Since, as we already mentioned, ${\bf R}$ is the vectorial displacement of
the hole after two consecutive nearest-neighbor hops, we find that
the hole band must have minima at $(\pm \pi/2,\pm\pi/2)$ 
%and $(0,\pm \pi)$ and $(\pm \pi,0)$ 
which is almost the case\cite{shraiman,elser,liu,boninsegni}.
We are now ready to discuss the question of why there is transfer of
spectral weight to higher string-excitations as we approach the 
$\Gamma$ point. As this point is approached
the kinetic energy lowering from the constructive interference
of strings differing by a segment of two overturned spins can
not be achieved through the phase factor associated  with the translation
operator (Bloch's phase factor) because ${\bf k} \to 0$. 
However, the desired phase coherence can be achieved
by forming quasiparticle states in which the various string states are not
included with amplitudes having the same phase but with an appropriate phase
difference so that the kinetic energy can take advantage from the 
overlap between two such string states. Therefore, although ${\bf k}\to (0,0)$
the quasiparticle state has {\it nodes} and, hence, 
it should  overlap more with higher string excitations.

A rather simplified physical picture of a mobile hole in a 
quantum antiferromagnet may be given. The hole becomes a well-defined
quasiparticle dressed with a cloud of strings of overturned spins
which the quasiparticle carries with it. The hole in this cloud of strings or 
``string-bag'' has
internal excitations, which, as we argued, have been possibly observed
in the high resolution ARPES studies\cite{ronning,graf}. We have shown
that the existence of such ``internal'' excitations of the ``spin-polaron''
are responsible for the vanishing of the quasiparticle from the
lowest string state at the $\Gamma$ point and for the transfer of
spectral weight to higher energy string excitations. In addition,
they are responsible for the recently reported high energy anomalous
dispersion in ARPES studies\cite{ronning,graf}.
%These excitations have been predicted and studied much before the ARPES 
%experiments.

%I wish to thank H. K. Ng for assisting me with Perl-GD programming
%in order to produce the intensity plots. 
This work was supported by NASA under grant No NAG-2867.


\begin{thebibliography}{99}
\bibitem{damascelli}A. Damascelli, Z. Hussain, Z. -X. Shen, Rev. Mod. Phys.
{\bf 75}, 473 (2003).
\bibitem{wells}B. O. Wells {\it et al.}, Phys. Rev. Lett. {\bf 74},
964 (1995).
\bibitem{RMP} E. Manousakis, Rev. Mod. Phys. {\bf 63}, 1 (1991).
%\bibitem{BR}W. Brinkman and T. Rice, Phys. Rev. {\bf B 2}, 1324 (1970).
\bibitem{liu} Z. Liu and E. Manousakis,  Phys. Rev. {\bf B 44}, 2414 (1991).
and $ibid$, {\bf  45},  2425 (1992). 
\bibitem{shraiman}
B. Shraiman and E. Siggia, Phys. Rev. Lett. {\bf 60}, 740 (1988);
{\bf 60}, 749 (1988); $ibid$ {\bf 60}, 740 (1988);
{\bf 61}, 467 (1988);
\bibitem{elser}
V. Elser, D. Huse, B. Shraiman, and E. Siggia, Phys. Rev. 
{\bf B 41}, 6715 (1990).
\bibitem{KLR} C. Kane, P. Lee, and N. Read, Phys. Rev. {\bf B 39}, 6880 (1989).
\bibitem{barnes}T. Barnes, {\it et al.}, Phys. Rev. B {\bf 40}, 10977 (1989).
\bibitem{dagotto}E. Dagotto, {\it et al.}, Phys. Rev. B {\bf 41}, 9049 (1990).
\bibitem{marsiglio}
F. Marsiglio, A. Ruckenstein, S. Schmitt-Rink and C. Varma,
Phys. Rev. {\bf B 43}, 10 882 (1991).
\bibitem{martinez}
G. Mart\'inez and P. Horsch,
Phys. Rev. {\bf B 44}, 317(1991).
\bibitem{boninsegni}
M. Boninsegni and E. Manousakis, Phys. Rev.  {\bf B 43}, 10 353 (1991).
$ibid$, {\bf B 45}, 4877-4884 (1992).
$ibid$, {\bf B 46}, 560-563 (1992).  
\bibitem{dagottor}E. Dagotto, Rev. Mod. Phys.,
{\bf 66}, 763 (1994) and references therein.
\bibitem{ZR}F. C. Zhang and T. M. Rice, Phys. Rev. {\bf B 37}, 3759 (1988).
\bibitem{kim}C. Kim {\it et al.}, Phys. Rev. Lett. {\bf 80}, 4245 (1998).
\bibitem{theory} E. Manousakis, in {\it Electronic, optoelectronic and
magnetic thin films}, ed. J. M. Marshall, N. Kirov and A. Vavrek,
pg 168, J. Wiley and Sons, (New York, 1995).
\bibitem{nazarenco}A. Nazarenko {\it et al.}, Phys. Rev. B {\bf 51}, 8676 
(1995). 
\bibitem{ferrell}B. Kyung and R. A. Ferrell, Phys. Rev. B {\bf 54}, 10125 
(1996). 
\bibitem{lee}T. K. Lee and C. T. Shih, Phys. Rev. B {\bf 55}, 5983 (1997).
\bibitem{wheatley}T. Xiang and J. M. Wheatley, Phys. Rev. B {\bf 54}, R12653 
(1996). 
\bibitem{belincher}V. I. Belinicher, {\it et al.},  Phys. Rev. B {\bf 54}, 14914 (1996).
\bibitem{eder}R. Eder, {\it et al.}, Phys. Rev. B {\bf 55}, 
R3414 (1997).
\bibitem{ronning}F. Ronning, {\it et al.}, Phys. Rev. B {\bf 71}, 
094518 (2005).
 \bibitem{graf}J. Graf, {\it el al.}, cond-mat/0607319.
%\bibitem{SVR}
%S. Schmitt-Rink, C. M. Varma, and A. E. Ruckenstein,
%Phys. Rev. Lett., {\bf 60}, 2793 (1988).
%\bibitem{TRUG}
%S. A. Trugman, Phys. Rev. {\bf B 37}, 1597 (1988).
%S. A. Trugman, $ibid$. {\bf B 41}, 892 (1990).
\end{thebibliography}
\end{document}